\documentstyle[12pt,epsf]{article}
\begin{document}

\input psfig

\title{Three Neutrino Flavors are Enough}

\author{Andy Acker and Sandip Pakvasa \\
Department of Physics and Astronomy\\
University of Hawai`i at Manoa, \\
Honolulu, HI, 96822 USA}

\date{}

\maketitle

\begin{abstract}
It is shown that it is possible to account for all three experimental
indications for neutrino oscillations with just three
neutrino flavors. In particular, we suggest that the solar and atmospheric
neutrino anomalies are to be explained by the same mass difference and mixing.
Possible implications and future tests
of the resulting mass-mixing pattern are given.
\end{abstract}

\section{Introduction}

\indent

Currently there are three pieces of evidence which
suggest that neutrinos have non-zero mass differences and
mixings. These are: (i) the observations of solar neutrinos,
(ii) the anomaly in the $\nu_{\mu}/\nu_e$ ratio in
atmospheric neutrinos at low energies and (iii) the possible
$\overline{\nu}_{\mu}-\overline{\nu}_{e}$ conversion
seen in the LSND experiment.
With the conventional interpretation of these effects
as being due to neutrino oscillations; the solar
neutrino anomaly needs a $\delta m^2$ $(\nu_e-\nu_x)$
of either about $10^{-5}-10^{-6} \; eV^2$ (MSW) or about
$10^{-10} \; eV^2$ (long wavelength vacuum oscillations), the
 atmospheric neutrino anomaly
calls for a $\delta m^2$ $(\nu_e-\nu_{\mu})$ or
$(\nu_{\mu}-\nu_{\tau})$ of $10^{-2}-10^{-3}\; eV^2$ and
the LSND effect needs a $\delta m^2$ $(\nu_e-\nu_{\mu})$
in the neighborhood of $1-2 \; eV^2$. For these three
independent $\delta m^2$'s at least one more neutrino
state (beyond the three flavors) is necessary \cite{Bilenky}.   

In this letter we explore the possibility
that all the neutrino anomalies may yet be accounted for
with just three flavors of neutrinos. We assume only two 
distinct values of $\delta m^2$'s. One value of 
$\delta m^2$ is selected to explain the low energy 
atmospheric data, while the second value is 
selected with the LSND effect in mind. 

Specifically we choose the following 
spectrum of $\delta m^2$'s:
\begin{equation} 
\delta m^2_{31} \sim \delta m^2_{32} \sim (1-2)eV^2
\end{equation}
\begin{equation} 
\delta m^2_{21} \sim10^{-2}\; eV^2
\end{equation}
We then seek to determine if, with this spectrum of 
$\delta m^2$'s, an explanation of the LSND, solar, atmospheric
neutrino data can be found by appropriate choice of
neutrino mixing angles.   

We begin by calculating the neutrino survival and 
transition probabilities. In general, these are given by
\begin{equation}
P_{\alpha\beta} =|\sum_{i} U_{\beta i} \exp(-iE_i t)U^*_{\alpha i}|^2
\label{genprob}
\end{equation}
Here the $U_{\alpha i}$ are elements of the matrix $U$ describing the mixing
between the flavor eigenstates ($\nu_{\alpha}$) and the mass eigenstates
($\nu_i$); that is $\nu_{\alpha}=\sum_iU_{\alpha i}\nu_i$. For
now we ignore possible CP violation, then $U$ is real, and Eq (\ref{genprob})  
may be written as 
\begin{equation}
P_{\alpha\beta}=\sum_i(U_{\beta i})^2(U_{\alpha i})^2
+2\sum_{i > j}U_{\beta i}U_{\beta j}U_{\alpha i} U_{\alpha j}
\times \cos \left(\frac{\delta m_{ij}^2 L}{2 E}\right)
\label{bprob}
\end{equation}
where $\delta m_{ij}^2 = m_i^2-m_j^2$ and $L$ is the distance between
the neutrino source and detection.
We present below an explicit form of the $3\times 3$ matrix $U$
\begin{equation}
U = \left( \begin{array}{ccc}
 C_{12}C_{13}  & C_{13}S_{12}&S_{13} \\
 - C_{23}S_{12}-S_{23}S_{13}C_{12} & 
C_{23}C_{12}-S_{23}S_{13}S_{12}& S_{23}C_{13}\\
   S_{23}S_{12}- C_{23}S_{13}C_{12} & -S_{23}C_{12}- 
C_{23}S_{13}S_{12} & C_{23}C_{13}\\
\end{array} \right)
\label{mat}
\end{equation}
where $C_{12}=\cos \theta_{12}$, $S_{12} = \sin \theta_{12}$, {\em etc.}.  
The explicit form of the transition probabilities 
depends on the spectrum of the $\delta m^2$'s.
For the choice of  $\delta m^2$'s considered here, all of  the oscillating 
terms in Eq  (\ref{bprob}) average to zero for the energies and path lengths 
relevant to both low energy atmospheric and solar neutrinos. Hence, for our model, the 
form of the  transition and survival probabilities relevant to 
solar and atmospheric neutrinos are: 
\begin{equation} 
P_{ee} =\sum_i (U_{ei})^4
\label{pee}
\end{equation}
\begin{equation} 
P_{\mu\mu} =\sum_i (U_{\mu i})^4
\label{pmm}
\end{equation}
\begin{equation} 
P_{e \mu}=P_{\mu e} =\sum_i (U_{ei}U_{\mu i})^2
\label{pme}
\end{equation}
Note that the above expressions are functions of the mixing angles only, and
are independent of the neutrino energy.

\section{Solar Neutrinos}

\indent

The four currently operating solar neutrino experiments
report the following results: 

\bigskip
\begin{tabular}{lcl}
\hline
{\em Experiment} &{\em Results} & \\ \hline
 Homestake\cite{cl}  &  $2.56 \pm 0.16\pm 0.14$ & SNU \\ \hline
 Kamioka \cite{kam}& $2.80 \pm 0.19 \pm 0.33$
 & $\times 10^{10}m^{-2}s^{-1}$ \\ \hline
 SAGE \cite{sage}  &  $72 \pm 12 \pm 7$ & SNU\\ \hline
 Gallex \cite{gallex}  &  $70 \pm 7 $ & SNU\\ \hline
 \end{tabular}
\bigskip

Recently measurements of the reaction 
$\gamma +{ ^{8}B}  \rightarrow { ^{7}Be} +p$ have been made\cite{reaction}. These
suggest that the cross-section for the inverse reaction
$^{7}Be +p \rightarrow {^{8}B} +\gamma$ at energies relevant
for the solar core may be somewhat smaller than the
value used in the Standard Solar Model(SSM) of Bahcall et al. calculations. Hence it is possible
that the flux of $^8B$ neutrinos is somewhat smaller than the
SSM, while the other neutrino fluxes are unaffected.
We allow for this possibility by defining $f_{B}$ as
\begin{equation}
f_{B}  = {\Phi_{B} \over \Phi^{BP}_{B}}
\end{equation}
where $\Phi^{BP}_{B}$ is the $^{8}B$ neutrino flux predicted 
in the SSM of Bahcall and Pinsonneault, which
incorporates helium diffusion\cite{BP}.
\footnote{Solar models which incorporate helium diffusion yield values for
the depth of the convective zone and primordial helium abundance which are
in excellent agreement with helioseismological data\cite{BP2}.}
Thus the parameter $f_{B}$ describes the deviation of 
the actual $^{8}B$ neutrino flux from the SSM value.

We can now proceed to describe the expected counting 
rates at the various solar neutrino experiments 
in terms of $f_{B}$ and $P_{ee}$ the solar $\nu_e$ survival 
probability.

With a threshold energy of  7.5 MeV the Kamiokande water Cerenkov detector is 
sensitive only to $^{8}B$ neutrinos. The expected flux is given by:
\begin{equation} 
R(\rm{KII}) =(P_{ee}+ {\alpha}
(1-P_{ee}))f_{B} \times 5.69\times 10^{10} m^{-2} s^{-1}
\end{equation}
Where $5.69 \times 10^{10} m^{-2}s^{-1}$ is the SSM prediction and 
$\alpha$ (approximately 0.16) is the ratio of the $\nu_{\mu (\tau)}-e$ to 
$\nu_{e}-e$ scattering cross sections integrated over the $^{8}B$ neutrino
spectrum. Note that we 
have ignored the possibility of oscillation into sterile flavors and 
assumed that solar neutrinos not interacting as $\nu_e$'s interact 
as either $\nu_{\mu}$'s or $\nu_{\tau}$'s with probability $1-P_{ee}$.

The expected counting rate in the Homestake $^{37}Cl$ experiment is given by
\begin{equation} 
R(^{37}Cl) = (6.2f_{B} + 1.8)\times P_{ee} \ \rm {SNU}
\end{equation}
where 6.2 SNU's is the expected contribution from $^{8}B$ 
neutrinos and 1.8 SNU's is the contribution from all other solar 
neutrino fluxes. Similarly, the expected counting 
rate in the  $^{71}Ga$ experiments SAGE and Gallex is given by
\begin{equation} 
R(^{71}Ga) = (13.8f_{B} + 117.6)\times P_{ee} \ \rm {SNU}
\end{equation}
where 13.8 SNU's is the expected contribution from $^{8}B$ neutrinos and 117.6 SNU's is
the contribution from all other solar neutrino fluxes.

The results of a chi-squared analysis are shown in Fig. 1.
We find that there is a solution at the 90\% C.L. when
the electron neutrino survival probability is in
the range $0.4< P_{ee} <0.55$ and the $^{8}B$ neutrino
flux is in the range $0.55< f_B <0.8$.  
\footnote{It was first suggested by Acker {\it et al.}\cite{A1} that
solar neutrinos might be accounted for by the same $\delta m^2$ and
mixing as atmospheric neutrinos and hence should show an energy
independent suppression. A search for an energy independent fit
with varying $f_B$ was made in Ref. \cite{KP}.}
This is consistent with the variation in $^{8}B$ neutrino flux found in an
analysis of solar models\cite{BP2}.

The results in Fig. 1 can be interpreted in terms of the neutrino
mixing matrix $U$. From Eq( \ref{pee}) the $\nu_e$ 
survival probability
$P_{ee}$ is a function of $\theta_{12}$ and $\theta_{13}$ only.
Each allowed value of $f_B$ in Fig. 1 corresponds to an allowed range of
$P_{ee}$ and hence to an allowed range of $\theta_{12}$ and 
$\theta_{13}$. In Fig. 2 we present a plot of the 
allowed values (90\% C.L.) of
$\sin(\theta_{12})$ and $\sin(\theta_{13})$ for 
$f_B = 0.8$ and  $f_B = 0.65$. At  $f_B = 0.8$,
$P_{ee}$ is required to be $\sim 0.43$, this can be realized only in three
flavor mixing. Hence, as shown in Fig. 2, $S_{12}$ and $S_{13}$
must both be nonzero.  At  $f_B = 0.65$, $P_{ee}$ 
can be greater then 0.5, which can be accomplished
in effective two flavor mixing. As shown in the figure, there
are allowed regions with  $\sin(\theta_{12})$ or $\sin(\theta_{13})$ 
equal to zero,
corresponding to pure $\nu_e - \nu_{\tau}$ or $\nu_e - \nu_{\mu}$
mixing respectively.

\section{LSND}

\indent

The Liquid Scintillation Neutrino Detector (LSND) experiment
at Los Alamos reports to have observed the possible appearance of
$\overline{\nu}_{e}$ in an initial beam of
$\overline{\nu}_{\mu}$'s \cite{LSND}. These results have been interpreted
as evidence of neutrino  oscillations and the preferred range of
$\delta m^2$
and $\sin^2(2\theta)$ in a two flavor mixing scenario given. For
definiteness, we choose $\sin^2(2\theta)\sim 1.2\times 10^{-3}$ and
$\delta m^2 \sim 2 \; eV^2$, which lie in this range \cite{Babu}. 

In the range of $L/E$ covered by the LSND set-up,
$\delta m_{12}^2L/4E \sim 0$ and 
$\delta m_{31}^2L/4E \sim \delta m_{32}^2L/4E$. Then
the $\overline{\nu}_{\mu} \rightarrow \overline{\nu}_{e}$ 
conversion probability is given by
\begin{equation}
P_{ \overline{\mu}  \overline{e}} = 4 U_{\mu 3}^2 U_{ e 3}^2
\sin^2(\delta m_{31}^2L/4E)
\end{equation}
 Thus the three-flavor interpretation of the LSND result is obtained by 
letting
\begin{equation}
\sin^2(2\theta_{LSND}) \rightarrow 4|U_{\mu 3}|^2|U_{e 3}|^2
\end{equation}
This may be expressed as a constraint on the three flavor mixing angles. Using
\begin{eqnarray}
|U_{e 3}|^2 =\sin^2(\theta_{13})
\nonumber\\
|U_{\mu 3}|^2 =\cos^2(\theta_{13})\sin^2(\theta_{23})
\end{eqnarray}
We obtain 
\begin{equation} 
\sin^2(\theta_{23})   = {\sin^2(2\theta_{LSND})\over 
4 \sin^2(\theta_{13}) \cos^2(\theta_{13})}
\end{equation}

Choosing
$\delta m_{31}^2 \sim \delta m_{32}^2$  to be near
$2 \; eV^2$, 
the LSND results then give $\sin^2(2\theta_{LSND})\sim 1.2\times 10^{-3}$, 
and we have 
\begin{equation} 
\sin(\theta_{23})^2 = {1.2\times 10^{-3} \over 4 \sin^2(\theta_{13})\cos^2(\theta_{13})}
\label{lsnd}
\end{equation} 

In order to determine the range of validity of this result,
constraints from reactor and accelerator experiments must
be taken into consideration. For this range of $\delta m^2$, reactor
experiments give the bound\cite{ue3}:
\begin{equation}
|U_{e3}|^2 \le 0.02
\end{equation}  
and accelerator experiments give the bound\cite{um3}:
\begin{equation}
|U_{\mu 3}|^2 \le 0.018.
\end{equation}  
As $|U_{\mu 3}|$ is related to $|U_{e 3}|$ through
Eq  (\ref{lsnd}) these two upper limits can be combined to form
bounds on the allowed values of $|U_{e3}|$ and $|U_{\mu 3}|$.
We have: 
\begin{equation}
0.129  \le |U_{e3}| \le 0.141
\label{limits}
\end{equation}
and
\begin{equation}
0.123  \le |U_{\mu 3}| \le 0.134
\end{equation} 
Hence the requirement that the LSND results be consistent with 
existing bounds on neutrino mixing leads to rather stringent limits on the 
allowed values of $|U_{e3}|$  and $|U_{\mu 3}|$. 
We will find these constraints particularly useful when 
interpreting  the atmospheric neutrino data.

\section{Atmospheric Neutrinos}

Experimentally measured atmospheric neutrino fluxes are often described
in terms of an (observed to predicted) 'ratio of ratios' $R$, where
\begin{equation}
R={(\nu_{\mu}/\nu_e)_{\rm{observed}}\over 
(\nu_{\mu}/\nu_e)_{\rm{Monte Carlo}}}
\end{equation}
 The final  results \cite{kam,klast} from Kamiokande for the low 
energy atmospheric neutrino $\nu_{\mu}/\nu_e$ ratio
place $R$ at $0.62 \pm  0.06\pm 0.06$. The results from IMB \cite{IMB} 
are in excellent agreement with these results. Results from non-water-Cerenkov
detectors are somewhat varying: Soudan \cite{Soudan} finds an 
R of $0.72 \pm 0.19 \; { ^{+0.05}_{-0.07}}$ 
whereas the results from Nusex \cite{Nussex} 
and Frejus \cite{Frejus} are consistent with an R of unity, although
with smaller statistics than the two large water-Cerenkov detectors.

It has been known for some time that this low energy atmospheric
neutrino anomaly can be explained by neutrino oscillations.
For $\delta m^2$ in the range 
$(4\times 10^{-3}-2\times 10^{-2}) \; eV^2$ it has been shown \cite{afit} 
that this anomaly can be explained by $\nu_{\mu}-\nu_{\tau}$
oscillations for
\begin{equation}
0.6 \le \sin^2(2\theta_{\mu-\tau}) \le 1.0
\label{mutauosc}
\end{equation}
or by $\nu_{\mu}-\nu_{e}$ oscillations for 
\begin{equation}
0.5 \le \sin^2(2\theta_{\mu-e}) \le 1.0.
\label{emuosc}
\end{equation}
Expressing these bounds in terms of the $U_{\alpha i}$ we have:
\begin{equation}
0.3 \le (P_{\mu \tau}=\sum_i (U_{\mu i}U_{\tau i})^2) \le 0.5
\label{pmto}
\end{equation}
for $\nu_{\mu}-\nu_{\tau}$ oscillations and
\begin{equation}
0.25 \le (P_{\mu e}=\sum_i (U_{\mu i}U_{e i})^2) \le 0.5
\end{equation}
for $\nu_{\mu}-\nu_{e}$ oscillations.

In the narrow range of $|U_{e3}|$ values permitted by
LSND, reactor and accelerator data (Eq \ref{limits})
$P_{\mu \tau}$ is less then 0.05 and thus inconsistent with Eq  (\ref{pmto}); 
while
$P_{\mu e}$ can take on values up to 0.48. Hence, in this
region, the atmospheric neutrino anomaly must be explained
almost exclusively by $\nu_{\mu}-\nu_{e}$ mixing. 

With $\theta_{23}$ constrained in terms of 
$\theta_{13}$ by Eq (\ref{lsnd}) 
and $\sin(\theta_{13})$ bound by Eq (\ref{limits}),
we find that $ 0.25 \le P_{\mu e} \le 0.5$ if
$\sin(\theta_{12})$ is in the range:
\begin{equation}
0.38 \le \sin(\theta_{12}) \le 0.92
\label{atmosc}
\end{equation}   
Thus we can express this explanation of the low energy 
atmospheric neutrino anomaly consistent with
the LSND, reactor, accelerator data as the region of the
$\sin(\theta_{12})-\sin(\theta_{13})$ plane bounded by
Eq (\ref{lsnd}) and Eq (\ref{limits}).

\section{A Combined Solution to the Solar, Atmospheric 
and LSND Data}

It is now a straightforward matter to identify 
simultaneous solutions to the Solar, Atmospheric
and LSND neutrino data. As the energy 
independent solution to the solar neutrino problem, 
shown in Fig. 2, and the combined LSND and 
atmospheric neutrino solution
are both expressed as regions of the
$\sin(\theta_{12})-\sin(\theta_{13})$ plane, any
intersection between the allowed regions represents 
the desired solution.

Fig. 3 presents a plot of
the intersecting regions of the Solar
neutrino and Atmospheric-LSND solutions.
Fig. 3a assumes the $^8B$ solar neutrino
flux, $f_{B}$, is at 80\% of its SSM value,
Fig. 3b assumes $f_{B}$ 
is at 70\% of its SSM value and
Fig. 3c assumes $f_{B}$ 
is at 65\% of its SSM value.
There is no intersection in Fig. 3a
and narrow region of overlap in
Fig. 3b broadening somewhat in Fig.
3c as the $^8B$ neutrino suppression is 
allowed to increase to 0.65.  

It should be noted that the 
selection of any region of the 
$\sin(\theta_{12})-\sin(\theta_{13})$ plane  
determines the complete
set of mixing angles, and hence
the neutrino mixing matrix $U$,
as $\sin(\theta_{23})$ is
fixed by Eq (\ref{lsnd}). Specifically the intersection
region of Fig. 3b corresponds to
$\sin(\theta_{12})$ $\sim$ 0.707,
$\sin(\theta_{13})$ $\sim$ 0.140 and
$\sin(\theta_{23})$ $\sim$ 0.125.

Using Eq (\ref{mat}) we present below the 
explicit form of the $3\times 3$
mixing matrix $U$ corresponding to
solution region of Fig. 3b:
\begin{equation}
U= \left( \begin{array}{ccc}
 .700 &  .700 & .140 \\
-.714 &  .689 & .124 \\
-.010 & -.187 & .982 \\
\end{array} \right)
\label{Ub}
\end{equation}
While for the solution region corresponding to Fig. 3c we
find that, in addition to Eq (\ref{Ub}) above,
the following range of matrix values are allowed:
\begin{equation}
 \left( \begin{array}{ccc}
 .630 &  .764 & .140 \\
-.776 &  .619 & .124 \\
-.010 & -.187 & .982 \\
\end{array} \right)
\leftrightarrow
 \left( \begin{array}{ccc}
 .764 &  .630 & .140 \\
-.645 &  .754 & .124 \\
-.028 & -.185 & .982 \\
\end{array} \right).
\label{Uc}
\end{equation}

\indent

\section{Implications}

\indent

{\bf (i)} Both Super-Kamiokande \cite{superk} and SNO \cite{SNO} should 
see {\bf NO} spectrum
distortion in either $\nu-e$ scattering or the
$\nu_e D$ charged current. The suppression in
$\nu-e$ scattering should be in the range 0.38 - 0.40 of SSM 
at all energies
and in $\nu_e D$, charged current suppression should be about 
$f_B P_{ee} \sim 0.32-0.34$.

{\bf (ii)}   Borexino \cite{borexino} should observe the 
$^{7}Be$ line at a rate of
$\left[ P_{ee}+\beta (1-P_{ee})\right]$ where $\beta$ is the ratio of
$\nu_{\mu (\tau)}-e$ to $\nu_{e}-e$ scattering cross sections. We thus expect
0.56 to 0.58 of the SSM rate, and an identical suppression should hold for
the pep line.

{\bf (iii)}  The atmospheric $\nu_\mu/\nu_e$ anomaly 
should be confirmed by Super-Kamiokande. Zenith angle dependence 
of multi-GeV neutrinos should confirm
the tentative evidence seen in Kamiokande \cite{afit}.  But most important is our
prediction \cite{A1} that $\nu_\mu- \nu_e$ oscillations should be confirmed by 
observation of excess high energy
e-like upcoming shower events (above and beyond $\nu_\mu$ neutral 
current events).  
  
{\bf (iv)} Future reactor experiments such as
CHOOZ \cite{chooz} and Palo Verde \cite {PV} which will be sensitive to
$\delta m^2$ upto $10^{-3} \; eV^2$ should see a
$\overline{\nu}_{e}$ survival probability of
$P_{ee}\sim 0.48-0.5$.

{\bf (v)} Long baseline experiments 
(such as MINOS \cite{minos}, CERN-LNGS \cite{cern} and KEK-PS E362
\cite{kek}) which will probe
$\delta m^2$ upto $10^{-3} \; eV^2$ should see
 $\nu_{\mu}-\nu_{\tau}$ conversion
with $P_{\mu \tau} =\sum_i (U_{\mu i}U_{\tau i})^2
\sim 0.028-0.035$ accompanied by 
$\nu_{\mu}-\nu_{e}$ conversion at $P_{\mu e} \sim 0.46-0.48$. 

{\bf (vi)} Short baseline experiments such as CHORUS \cite{nomad},
NOMAD \cite{nomad} and COSMOS \cite{cosmos} will 
probe $\nu_{\mu}-\nu_{\tau}$ conversion for $\delta m^2 \ge 0.1 \; eV^2$.
We predict, at $\delta m^2 = 1-2 \; eV^2$, an effective
$\sin^2(2\theta)$ of $4(U_{\mu 3}U_{\tau 3})^2$ which is 0.06.

{\bf (vii)} Large mixings in some ranges of $\delta m^2$
lead to strong conversion of $\overline{\nu}_{\mu}$ to
$\overline{\nu}_{\tau}$ due to the MSW effect in
supernova, leading to a harder energy spectrum of the
emerging $\overline{\nu}_e$'s. This can lead to
potential conflict with observation of neutrinos
from SN1987A. For the $\delta m^2$ in our scenario this
is not a problem \cite{SN}.

{\bf (viii)} The neutrino mass spectrum implied by
our scenario is:
\begin{eqnarray}
m_1 \sim m_0
\nonumber\\
m_2 \sim m_0 + \epsilon
\nonumber\\
m_3 \sim \sqrt{m_{2}^2 + 2 eV^2}
\end{eqnarray}
where $\delta m_{12}^2 \sim (2m_0\epsilon + \epsilon^2)
\sim 10^{-2} \; eV^2$. There are two limiting cases of
interest assuming that the largest mass is in the $eV$ range.
One is the hierarchical limit, in which
$m_0$ is negligible. Then $m_1 \ll m_2 \sim 0.1 eV$
and $m_3 \sim 1.4  eV$. The other is the nearly 
degenerate limit, in which
\begin{eqnarray}
m_1 \sim 1  eV
\nonumber\\
m_2 \sim (1 + \epsilon)  eV
\nonumber\\
m_3 \sim 1.73  eV
\end{eqnarray}
with $\epsilon \sim {1\over 2}(10^{-2})eV$. Then,
 the sum of the neutrino masses is
$\sum_i m_i \sim 4 eV$. In this case, the
Cosmological density parameter associated with
neutrinos $\Omega_{\nu} =
0.011 h^{-2}\sum_i m_{i} = 0.044 h^{-2} \approx 0.2$ (for h of about 0.5) and 
the amount of neutrino dark matter component along with cold dark matter
makes for a viable and testable scenario for mixed dark matter \cite{DM}.

{\bf (ix)}  When the neutrinos are Majorana particles, the effective
mass \newline 
$<m_{\nu_e}>$ relevant in neutrino-less
double $\beta$-decay analysis is
\begin{equation}
<m_{\nu_e}> = \sum_i U^2_{ei}m_i
\end{equation}
We find that in the case of the hierarchical spectrum
$<m_{\nu_e}> \sim 0.1 eV$ whereas in the degenerate
case $<m_{\nu_e}> \sim 1 eV$ (this could be somewhat
smaller when CP phases are taken into account). It is interesting that these
values are in the range of what the double beta decay experiments can probe
now and in the near future \cite{cp}.

{\bf (x)}  When the mixing matrix is allowed to have a CP violating phase,
the CP violating neutrino flavor conversion probability differences are
given by \cite{meff}  

\begin{eqnarray}
\Delta P = P_{\mu \tau} - P_{\bar{\mu} \bar{\tau}} = P_{\bar{\mu} \bar{e}} 
- P_{\mu e}
\nonumber\\
= -4 J_{cp}^\nu \left [ \sin D_{12} + \sin D_{23} + \sin D_{31} \right ]
\end{eqnarray}
where
\begin{eqnarray}	
J_{CP}^\nu = Im \left [ U_{\mu 2} U_{\tau 2}^* U_{\mu 3}U_{\tau 3}^* \right ] 
\nonumber\\
= |U_{\mu 2}||U_{\tau 2}||U_{\mu 3}||U_{\tau 3}|\sin \phi,
\end{eqnarray}
and
\begin{equation}
 D_{ij} = \delta m^2_{i j} L/2E
\end{equation}
with  $\phi$ being the phase in the mixing matrix.  With the
matrix of Eq.( {\ref{Uc}}), $J_{CP}^\nu \leq 0.07;$ and $\left[ \sin D_{12 }+ \sin D_{23} +
\sin D_{31}\right]
\approx \sin D_{12}$ is given by $\sim -1$ for 
$L/E =730$km/10GeV (relevant for MINOS) and also for $L/E=250$km/3GeV 
(relevant for E362).  
Hence, $\Delta P$ can be as large as 0.07.(For these parameters, matter
effects are negligible \cite{meff}).

We conclude by stressing that our proposal to account for both solar and
atmospheric neutrino anomalies by the same mass and mixing can be confirmed
or ruled out in the very near future.

\section*{Acknowledgements}

\bigskip

We thank V. Barger, A. Joshipura, L. Kofman, J.G. Learned, S. Parke,
R. S. Raghavan, H. Sugawara, X. Tata and T.J. Weiler for
valuable discussions and encouragement.
This research was supported in part by the  U.S. 
Department of Energy grant \#DE-FG-03-94ER40833.

\bigskip

\vfill \eject

\vfill \eject

\bigskip

\section*{Figure Captions}

\bigskip

\indent

{\bf Figure 1} Contour plot showing the allowed  
values of the parameters $P_{ee}$
and $ f_{B}$n at the 90 \% (solid line) and 95 \% (dashed line) confidence 
levels in the three flavor mixing solution to the solar 
neutrino problem.

\bigskip

{\bf Figure 2} Contour plot showing the allowed values
of  $\sin(\theta_{12})$ and $\sin(\theta_{13})$  
(90 \% confidence level) in the three flavor mixing
solution to the solar neutrino problem for fixed values of $f_B$. 
Solid line; $f_B=0.8$, dashed line; $f_B=0.65$.

\bigskip

{\bf Figure 3} Combined solution: solar atmospheric and
LSND results. Dashed line; bounds from LSND, reactor and accelerator
experiments, Hatched region; Atmospheric neutrino
anomaly explained by $\nu_{\mu}-\nu_e$ oscillations,
Solid line; allowed region at the 90 \% C.L.
in the three flavor mixing solution to the solar neutrino
problem for: (a)  $f_B=0.8$ (b) $f_B = 0.7$ and (c) $f_B=0.65$.

\vfill  \eject

\begin{figure}[htb]
\centerline{\epsfysize 8.5 truein
\epsfbox{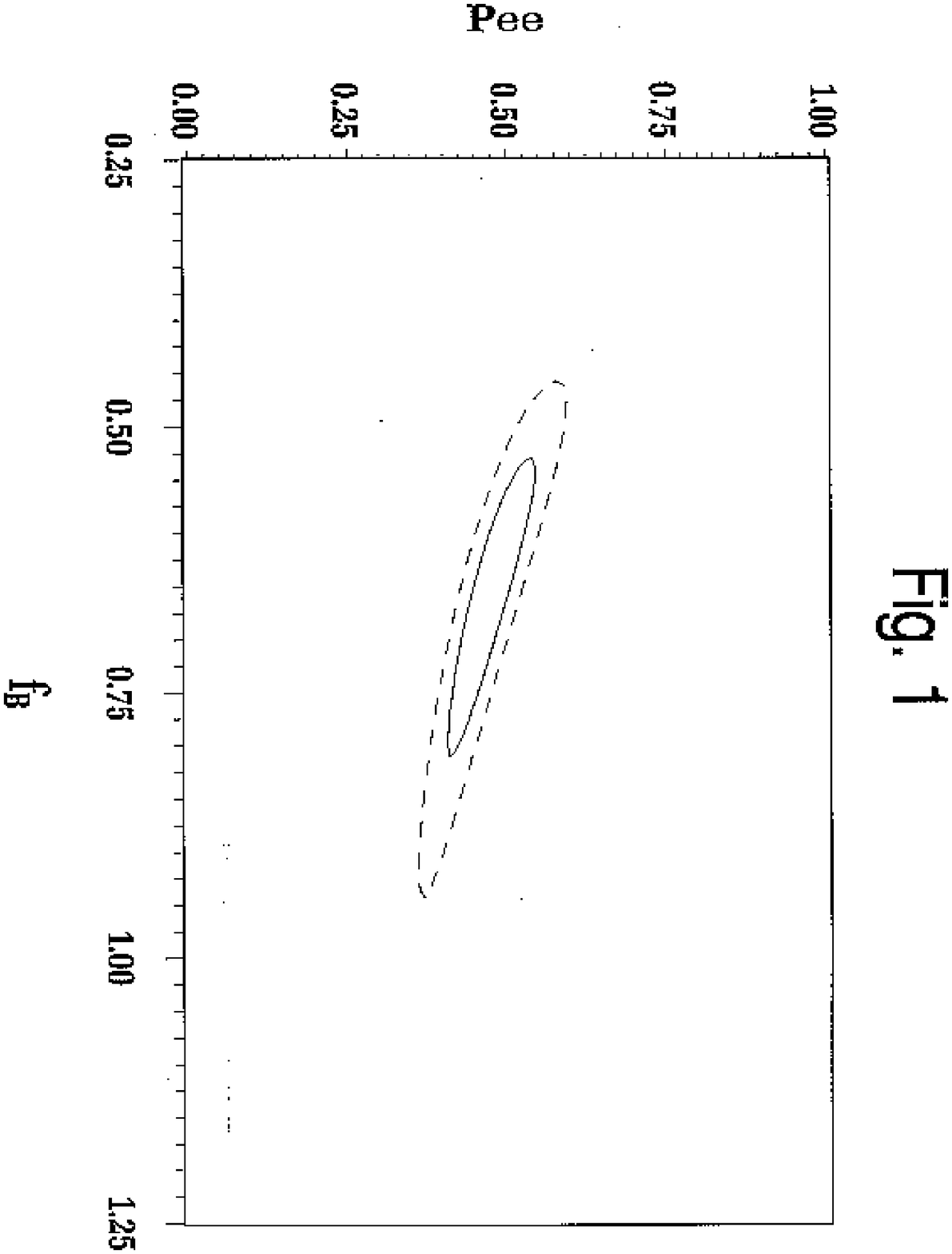}}
\end{figure}
\vfill   \eject

\begin{figure}[htb]
\centerline{\epsfysize 8.5 truein
\epsfbox{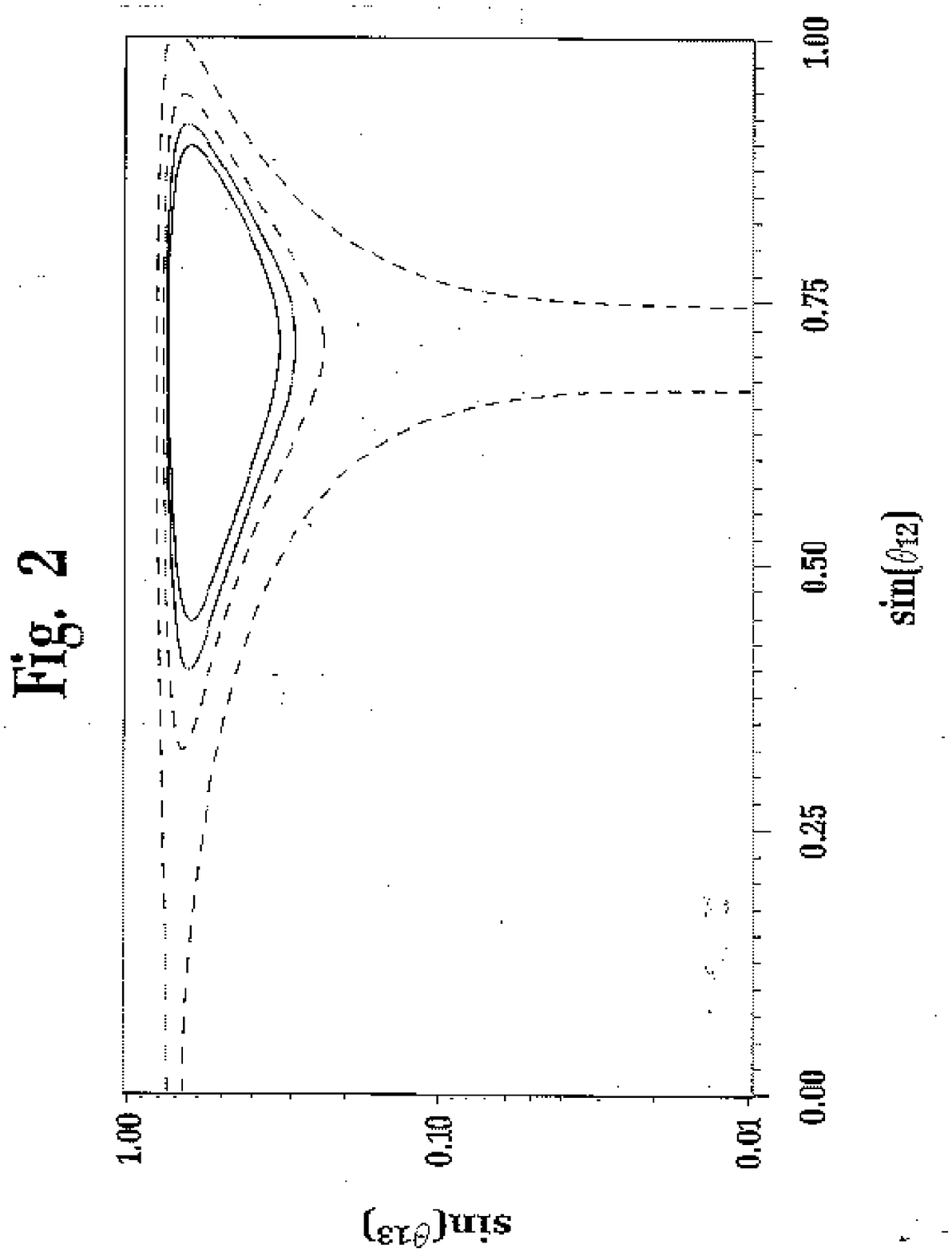}}
\end{figure}
\vfill   \eject

\begin{figure}[htb]
\centerline{\epsfysize 8.5 truein
\epsfbox{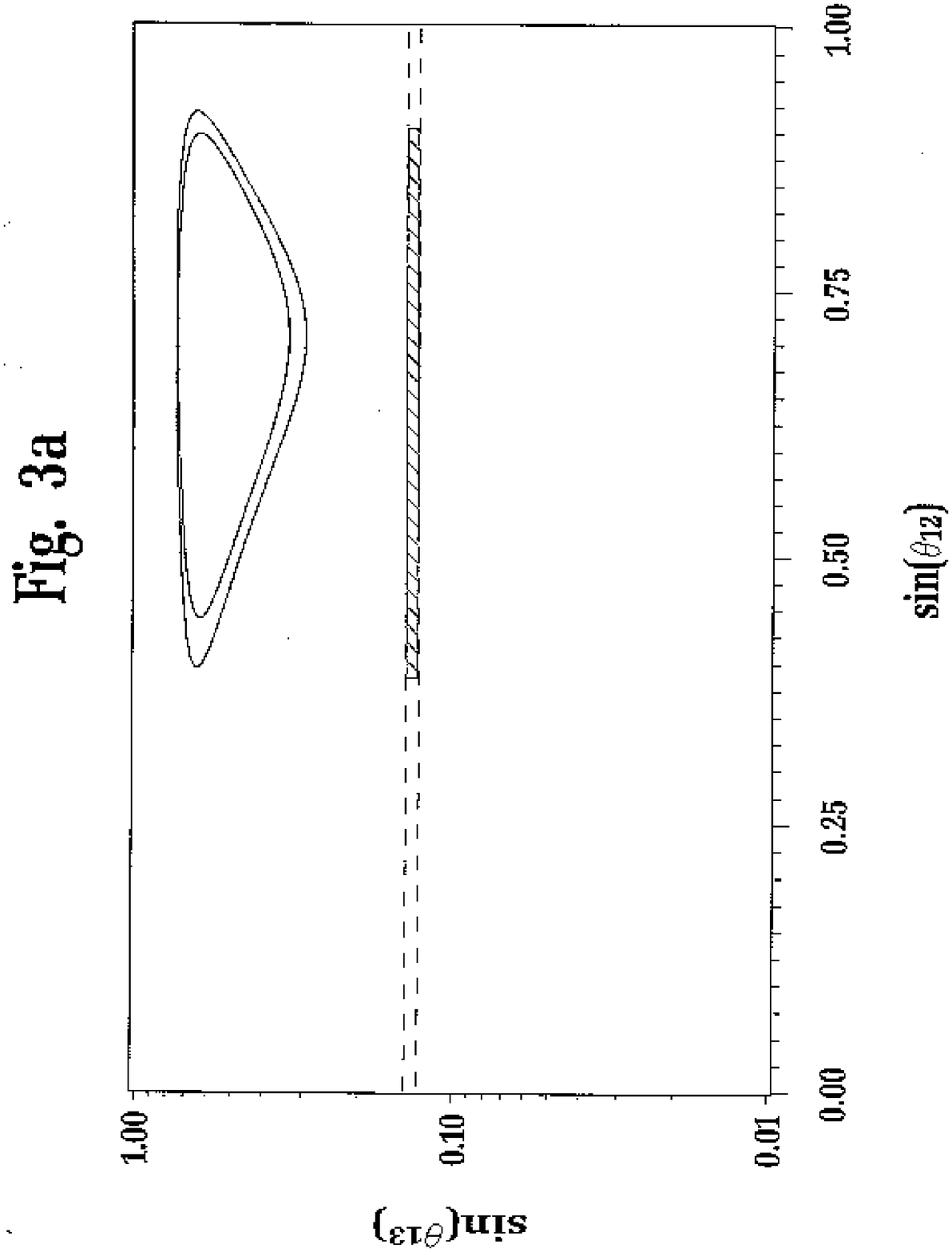}}
\end{figure}
\vfill   \eject

\begin{figure}[htb]
\centerline{\epsfysize 8.5 truein
\epsfbox{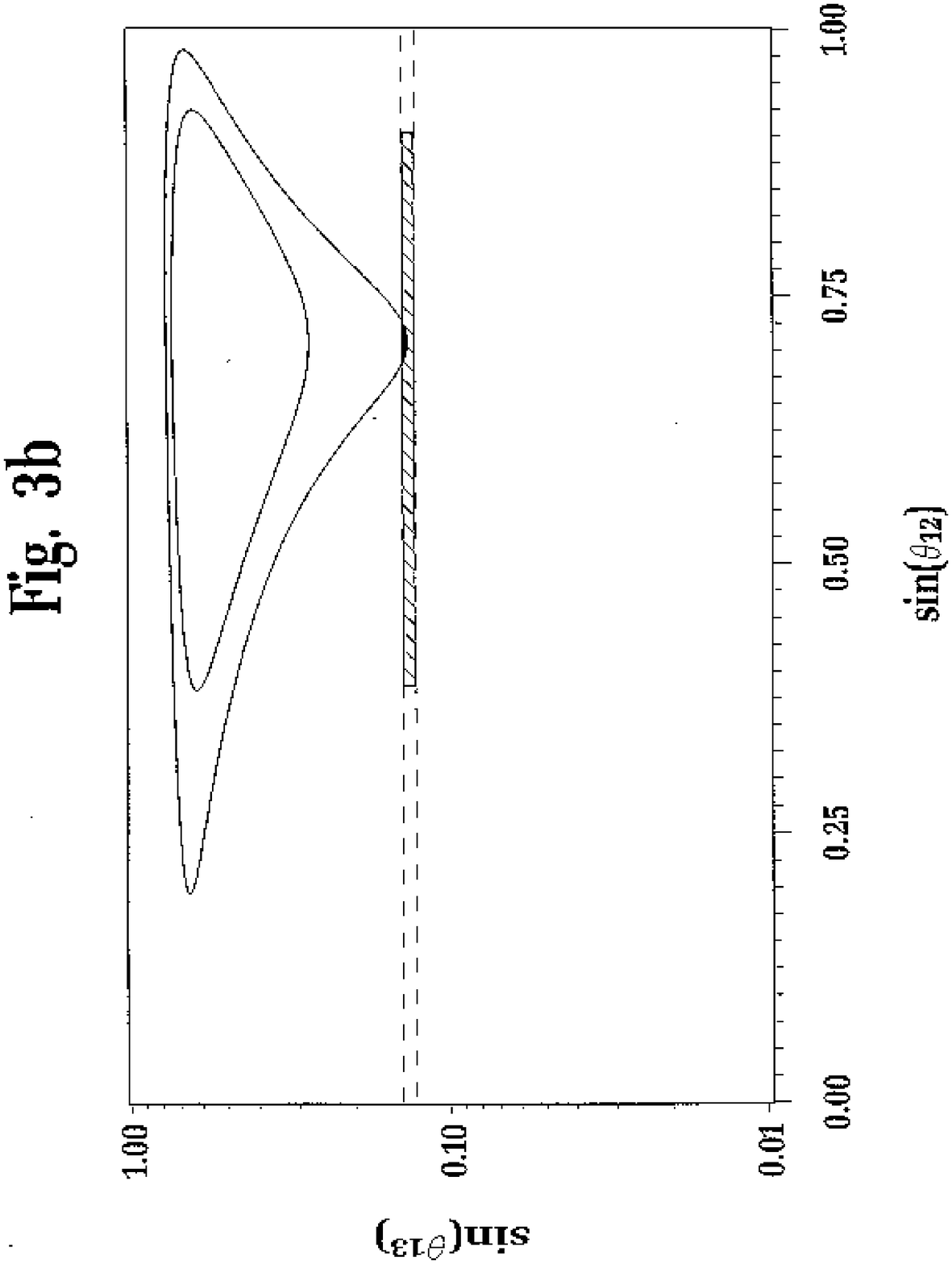}}
\end{figure}
\vfill   \eject

\begin{figure}[htb]
\centerline{\epsfysize 8.5 truein
\epsfbox{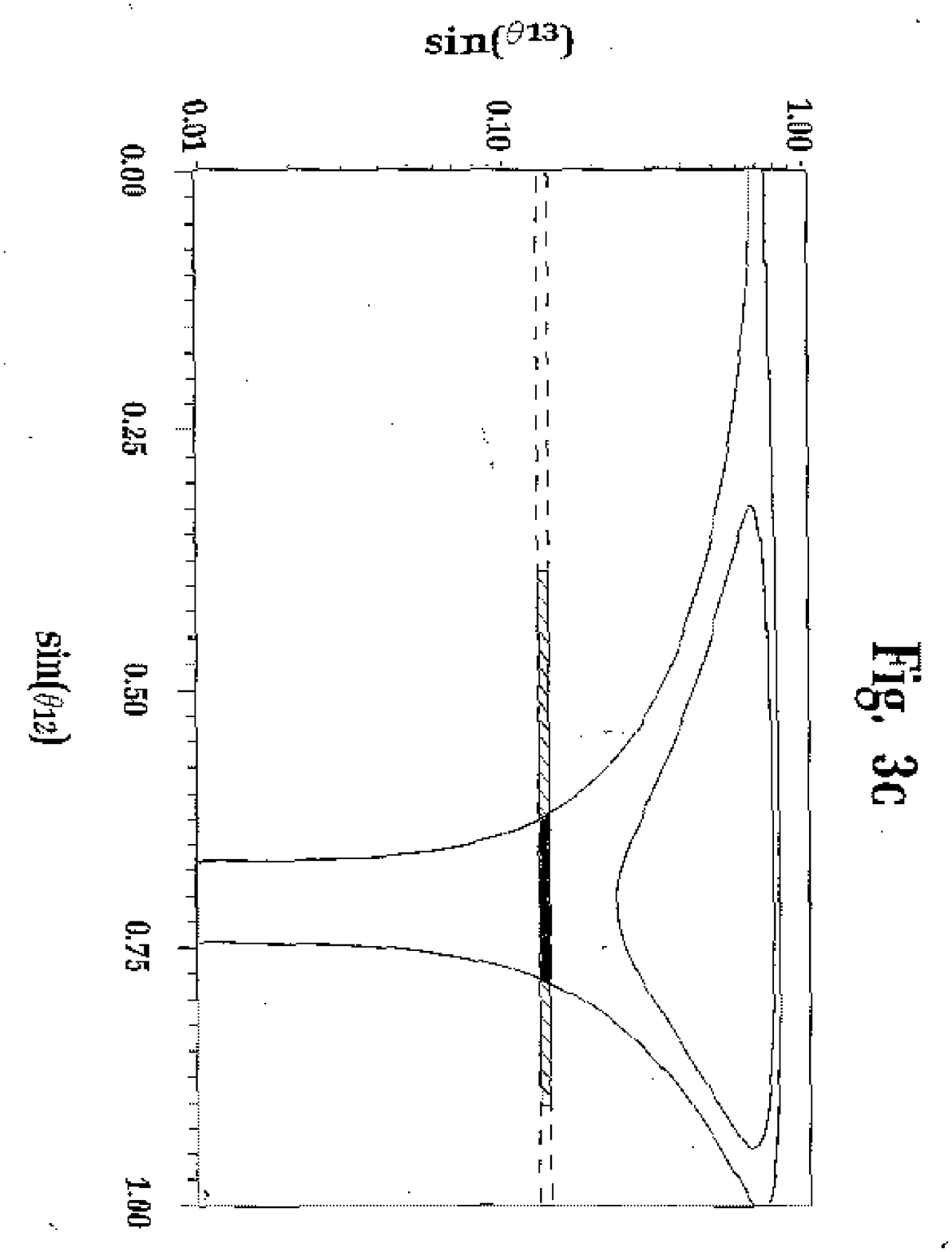}}
\end{figure}
\vfill    \eject

\end{document}